\documentclass[a4,12pt]{article}
\usepackage{type1cm,amsmath,amssymb,color,graphics,amscd,amsfonts,epsf,indentfirst,cite}
\usepackage{epsfig}
\usepackage{subfigure}
\usepackage[dvipdfm,hypertex]{hyperref}
\makeatletter
\@addtoreset{equation}{section}

\makeatletter
\renewcommand\section{\@startsection {section}{1}{\z@}%
                                   {-3.5ex \@plus -1ex \@minus -.2ex}%
                                   {2.3ex \@plus.2ex}%
                                   {\normalfont\large\bfseries}}
\renewcommand\subsection{\@startsection{subsection}{2}{\z@}%
                                     {-3.25ex\@plus -1ex \@minus -.2ex}%
                                     {1.5ex \@plus .2ex}%
                                    {\normalfont\bfseries}}

\def\no{\nonumber \\}
\def\btab{\begin{table}[h] \begin{center} \begin{tabular}{l lp{3in}}}
      \def\etab{\end{tabular} \end{center} \end{table}}
\def\btabm{\begin{center} \begin{tabular}}
    \def\etabm{\end{tabular} \end{center}}

\def\ie{{\it i.e.}}

\def\a{{\alpha}}

\def\g{{\gamma}}
\def\G{{\Gamma}}

\def\t{{\theta}}

\def\f#1#2{{\frac{#1}{#2}}}

\def\s{\sqrt}

\def\f {\frac}
\def\ti{\tilde}

\def\p{\partial}
\def\we{\wedge}

\newcommand{\half}{{1\over 2}}

\begin{document}

\begin{titlepage}
  \thispagestyle{empty}

  \begin{flushright}
    KUNS-2167\\
  \end{flushright}

  \vspace{2cm}

  \begin{center}
    \font\titlerm=cmr10 scaled\magstep4
    \font\titlei=cmmi10 scaled\magstep4
    \font\titleis=cmmi7 scaled\magstep4
     \centerline{\titlerm
      CFT Duals for Extreme Black Holes}

    \vspace{1.5cm}
    \noindent{{
        Thomas Hartman$^\dag$\footnote{e-mail: hartman@physics.harvard.edu},
        Keiju Murata$^\S$\footnote{e-mail: murata@tap.scphys.kyoto-u.ac.jp},
        Tatsuma Nishioka$^\S$\footnote{e-mail: nishioka@gauge.scphys.kyoto-u.ac.jp},
        and Andrew Strominger$^\dag$\footnote{e-mail: andy@physics.harvard.edu}
      }}\\
    \vspace{0.8cm}

   {\it $^\dag$Center for the Fundamental Laws of Nature\\
    Jefferson Physical Laboratory, Harvard University, Cambridge, MA 02138, USA}\\

    \vspace{0.5cm}

   {\it $^\S$Department of Physics, Kyoto University, Kyoto 606-8502, Japan} \\
   \vspace{1cm}
   {\large \today}
  \end{center}

  \vskip 2em

  \begin{abstract}
    It is argued that the general four-dimensional extremal Kerr-Newman-AdS-dS black hole is holographically dual to a (chiral half of a) two-dimensional CFT, generalizing an argument given recently for the special case of extremal  Kerr. Specifically, the asymptotic symmetries of the near-horizon region of the general extremal black hole are shown to be generated by a  Virasoro algebra.
    Semiclassical formulae are derived for the central charge and temperature of the dual CFT as functions of the cosmological constant, Newton's constant and the black hole charges and spin.  We then show, assuming the Cardy formula,  that the microscopic entropy of the dual CFT precisely
    reproduces the macroscopic Bekenstein-Hawking area law.
This CFT description becomes singular in the extreme Reissner-Nordstrom limit where the black hole has no spin.  At this point a second  dual
CFT description is proposed in which the global part of the $U(1)$ gauge symmetry is promoted to  a Virasoro algebra.  This second description is also found to reproduce the area law. Various further generalizations including higher dimensions are discussed.
  \end{abstract}

\end{titlepage}

\tableofcontents

\section{Introduction}
The origin of the Bekenstein-Hawking area law for black hole entropy remains to be fully understood.   The microscopic origin of the entropy of certain extremal supersymmetric black holes in string theory was explained in  \cite{SV} by counting BPS states.  While supersymmetry seemed to play a crucial role in that discussion, recent investigations of the attractor
mechanism (see \cite{Sen} and references therein) indicate that really only extremality is needed.  Recently, a new duality called the Kerr/CFT correspondence
between four-dimensional non-supersymmetric but extremal Kerr black holes and a two-dimensional CFT was proposed \cite{GHSS},
and the Bekenstein-Hawking entropy of the black holes was reproduced as the statistical entropy of the
dual CFT using the Cardy formula. Here extremality is once again important in order to take the near
horizon limit of the black hole and define the dual CFT.\footnote{See
\cite{Car,Car2,Park,KKP} for earlier work, and \cite{HHKNT} for some recent progress.}

In this paper we show that the Kerr/CFT correspondence generalizes to the Kerr-Newman-AdS-dS/CFT correspondence  in four dimensions.  The near horizon metric of these general extremal black holes has the $U(1)\times SL(2,R)$ symmetric form\footnote{In fact, it was proved in \cite{KLR,KuLu} that a very general class of extremal black holes  has a near horizon metric of this form.}
\begin{equation}\label{GenExt}
 ds^2=\Gamma(\theta)\left[
-r^2dt^2+\frac{dr^2}{r^2}
+\alpha (\t) d\theta^2 \right] + \gamma(\theta)(d\phi+krdt)^2\ .
\end{equation}
We argue that quantum gravity on this geometry is dual to a 2D CFT and   derive the central charge. We further compute the temperature  and, assuming the Cardy formula, show that the statistical entropy of the CFT is exactly equal to the Bekenstein-Hawking entropy.\footnote{It is suggested in \cite{ANT} that using \cite{RyuTa}, the extremal black hole entropy of (\ref{GenExt}) can also be related to the entanglement entropy between two CFTs dual to AdS$_2$, including higher derivative
corrections.}

The calculation of the central charge is similar to \cite{GHSS}, which was based on the approach originally taken by Brown and
Henneaux for AdS$_3$ \cite{brownhenneaux}. We impose boundary conditions on the metric identical to those for near horizon extreme
Kerr, and show that the asymptotic symmetries form a Virasoro algebra with central charge\footnote{Here $a=J/M_{ADM}$ with $M_{ADM} = \frac{r_+[(1+r_+^2/\ell^2)^2-q^2/\ell^2]}{(1-r_+^2/\ell^2)(1-a^2/\ell^2)^2}$,  $q^2$ is related
to the usual electric and magnetic charges by $q^2 = (1-a^2/\ell^2)^2(Q_e^2 + Q_m^2)$ and the horizon radius $r_+$ is the solution of  $a^2 = \frac{r_+^2(1+3r_+^2/\ell^2)-q^2}{1-r_+^2/\ell^2}$.}
\begin{align}\label{GenCC}
 c_L &=   3k \int_0^\pi d\t \s{\G (\t) \a (\t) \g (\t)} \\
    &=\frac{12r_+\sqrt{(3r_+^4/\ell^2 + r_+^2 - q^2)(1-r_+^2/\ell^2 )}}{1+6r_+^2/\ell^2 -3r_+^4/\ell^4 -q^2/\ell^2 }\ .
\end{align}

The ``temperature" of the CFT, \ie \,the thermodynamic potential dual to the zero mode of the Virasoro algebra,  follows from the first law of black hole thermodynamics, and works out to\begin{align}\label{CFTTL}
    T_L = \f{1}{2\pi k}\  .
\end{align}
Plugging this into the Cardy formula we obtain the statistical entropy
\begin{align}\label{CFTent}
    S = \f{\pi^2}{3} c_L T_L = \f{\pi }{2}\int_0^\pi d\t \s{\G (\t) \a(\t) \g (\t)} \ ,
\end{align}
which is equal to the Bekenstein-Hawking entropy of (\ref{GenExt}).

The central charge of the Kerr-Newman-AdS-dS black hole is proportional to angular
momentum $J$, so in the Reissner-Nordstrom-AdS limit $J\rightarrow 0$, the central charge vanishes.
In the same limit, the temperature blows up, so the Cardy formula still produces the correct entropy, but the description is clearly breaking down. In this situation we should seek an alternate well-behaved dual CFT description. The existence of more than one CFT description is to be expected. For example, the 5D black hole of  \cite{SV} has multiple dual CFT descriptions. The central charge of the dual CFT is determined by two of the three charges $Q_1,~~Q_5$ and $n$, but which two (or combination thereof) depends on the duality frame.
The alternative description we employ requires the additional assumption (discussed in section 7) that we can treat the total space of the 4D geometry plus the $U(1)$ gauge bundle as a 5D geometry in its own right. When $J=0$, the boundary conditions adopted for nonzero $J$  degenerate, and there are alternate boundary conditions which extend the  global $U(1)_{gauge}$ to a Virasoro. With vanishing cosmological constant, the central charge is then found with the usual methods to be
\begin{equation}
c = 6 Q_e(Q_e^2 + Q_m^2) \ ,
\end{equation}
where $Q_e$ and $Q_m$ are the electric and magnetic charges of the black hole, and the associated temperature is
\begin{equation}
T = {1\over 2\pi Q_e}\ .
\end{equation}
Together these reproduce the Bekenstein-Hawking area law via the Cardy formula.

It seems quite likely that the derivation of the entropy (\ref{CFTent}) is  applicable to the very general class of black holes and near horizon geometries of the form  (\ref{GenExt}) found in \cite{KLR,KuLu}, including the Kaluza-Klein reduction of higher dimensional black holes, but we have not checked all the details. We will discuss this issue in section 8.

The organization of this paper is as follows. In section 2 we review the geometry of the Kerr-Newman-AdS-dS black hole and
its near horizon limit. We find the asymptotic symmetry group for a general class of black holes in Einstein-Maxwell theory
in section 3, and compute the central charge in section 4.  The temperature is computed in section 5 and used in conjunction
with the central charge to derive the microscopic entropy in section 6.  In section 7 we consider Reissner-Nordstrom-AdS black holes,
and finally we comment on a more general class of black holes in section 8.

While this paper was in preparation, \cite{LMP} appeared, which generalizes the Kerr/CFT correspondence to Kerr-AdS black holes
in four and higher dimensions. After this paper appeared on the ArXiv, we received \cite{AOT}, which treats the 5D rotating Kaluza-Klein black holes.
These have some overlap with our results.

\section{Kerr-Newman-AdS-dS Black Holes}\label{s:KNA}
\subsection{Geometry}
In this section we review the four-dimensional Kerr-Newman-AdS-dS black hole.  It is a solution of the Einstein-Maxwell action
\begin{equation}\label{einsteinmaxwell}
S = {1\over 16 \pi}\int d^4x \sqrt{-g}\left(R + {6\over \ell^2}-{1\over 4}F^2\right)\ ,
\end{equation}
where $\ell^2$ is positive (negative) for AdS (dS).
In both cases the metric is (eg, \cite{CCK})
\begin{equation}\label{KNAdS}
 ds^2=-\frac{\Delta_r}{\rho^2}\left(d\hat{t}-\frac{a}{\Xi}\sin^2\theta d\hat{\phi}\right)^2
+\frac{\rho^2}{\Delta_r}d\hat{r}^2 + \frac{\rho^2}{\Delta_\theta}d\theta^2
+\frac{\Delta_\theta}{\rho^2}\sin^2\theta\left(ad\hat{t}-\frac{\hat{r}^2+a^2}{\Xi}d\hat{\phi}\right)^2\ ,
\end{equation}
with
\begin{equation}
\begin{split}
&\Delta_r=(\hat{r}^2+a^2)\left(1+\frac{\hat{r}^2}{\ell^2}\right)-2M\hat{r}+q^2\ , \qquad \Delta_\theta=1-\frac{a^2}{\ell^2}\cos^2\theta\ ,\\
&\rho^2=\hat{r}^2+a^2\cos^2\theta\ ,\qquad \Xi=1-\frac{a^2}{\ell^2}\ ,\qquad q^2=q_e^2+q_m^2\ .
\end{split}
\end{equation}
The horizons are located at the zeros of $\Delta_r$.  We denote the value of $\hat{r}$ at the outer horizon by $r_+$. The gauge field and field strength are
\begin{align}\label{FS}
    A &= - \f{q_e \hat{r}}{\rho^2}\left( d\hat{t} - \f{a\sin^2\t}{\Xi}d\hat{\phi} \right)
        - \f{q_m\cos\t}{\rho^2}\left( ad\hat{t} - \f{\hat{r}^2 + a^2}{\Xi}d\hat{\phi} \right) , \\
    F &= - \f{q_e(\hat{r}^2 - a^2\cos^2\t) + 2q_m \hat{r}a\cos\t}{\rho^4}
    \left( d\hat{t} - \f{a\sin^2\t}{\Xi}d\hat{\phi} \right) \we d\hat{r} \no
    & \quad + \f{q_m(\hat{r}^2 - a^2\cos^2\t) - 2q_e\hat{r}a\cos\t}{\rho^4}\sin\t d\t \we
    \left( ad\hat{t} - \f{\hat{r}^2 + a^2}{\Xi} d\hat{\phi}  \right).
\end{align}
The angular velocity of the horizon and the entropy are
\begin{align}\label{OhS}
    &\Omega_H = \f{\Xi a}{(r_+^2 + a^2)},\qquad S = {\mbox{Area}\over 4} = \pi \f{r_+^2 + a^2}{\Xi},\\
    &\Omega_H^\infty=\Omega_H+\frac{a}{\ell^2}=\frac{a(1+r_+^2/\ell^2)}{r_+^2+a^2}\ ,
\end{align}
where $\Omega_H^\infty$ is the angular velocity measured at spatial infinity, which is used in certain thermodynamic relations.
The Hawking temperature is
\begin{align}\label{THKN}
T_H = \frac{r_+(1+a^2/\ell^2+3r_+^2/\ell^2-(a^2+q^2)/r_+^2)}
{4\pi(r_+^2+a^2)}\ .
\end{align}
The physical mass, angular momentum, and electric and magnetic charges are
\begin{equation}\label{Charges}
M_\text{ADM}=\frac{M}{\Xi^2}\ ,\quad
J=\frac{aM}{\Xi^2}\ ,\quad
Q_e=\frac{q_e}{\Xi}\ ,\quad
Q_m=\frac{q_m}{\Xi}\ .
\end{equation}

\subsection{Extreme Limit}
In the extreme limit, the inner and outer horizons degenerate to a single horizon at $r_+$.
The extremality condition is
\begin{equation}
\begin{split}
&a^2 = \frac{r_+^2(1+3r_+^2/\ell^2)-q^2}{1-r_+^2/\ell^2}\ ,\\
&M = \frac{r_+[(1+r_+^2/\ell^2)^2-q^2/\ell^2]}{1-r_+^2/\ell^2}\ ,
\end{split}
\end{equation}
and the entropy at extremality is
\begin{align}\label{TE}
S(T_H=0) = \f{\pi(2r_+^4/\ell^2+2r_+^2-q^2)}{1-2r_+^2/\ell^2-3r_+^4/\ell^4+q^2/\ell^2}.
\end{align}

\subsection{Near Horizon Limit}
To find the near horizon geometry of the extreme Kerr-Newman-AdS-dS black hole,
we introduce new coordinates following \cite{Bardeen:1999px}
\begin{equation}\label{NearHor}
\begin{split}
&\hat{r}=r_+ + \epsilon r_0 r\ ,\\
&\hat{t}=t r_0/\epsilon\ ,\\
&\hat{\phi}=\phi+\Omega_H\frac{t r_0}{\epsilon}\ ,
\end{split}
\end{equation}
where $r_0$ is defined below, and take the limit $\epsilon\rightarrow 0$. The metric becomes
\begin{equation}\label{NHKNA}
 ds^2=\Gamma(\theta)\left[
-r^2dt^2+\frac{dr^2}{r^2}
+\alpha (\t) d\theta^2 \right] + \gamma(\theta)(d\phi+krdt)^2\ ,
\end{equation}
where
\begin{eqnarray}\label{definefuncs}
\Gamma(\theta) &=& \frac{\rho_+^2 r_0^2}{r_+^2+a^2}\ , \no
\alpha(\theta) &=& \frac{r_+^2+a^2}{\Delta_\theta r_0^2}\ , \\
\gamma(\theta) &=& \frac{\Delta_\theta(r_+^2+a^2)^2\sin^2\theta}{\rho_+^2\Xi^2}\ ,\notag
\end{eqnarray}
and we have defined
\begin{equation}\label{defineks}
\rho_+^2=r_+^2+a^2\cos^2\theta\ ,\quad
 r_0^2=\frac{(r_+^2+a^2)(1-r_+^2/\ell^2)}{1+6r_+^2/\ell^2-3r_+^4/\ell^4-q^2/\ell^2}\ ,\quad
k = \frac{2ar_+\Xi r_0^2}{(r_+^2+a^2)^2}\ .
\end{equation}
The field strength (\ref{FS}) becomes
\begin{align}
    F = f(\t) kdr\we dt + f'(\t) (d\t\we d\phi + krd\t \we dt) \ ,
\end{align}
and the near horizon gauge field is
\begin{equation}\label{nearhorizonA}
A = f(\theta)(d\phi + k r dt)
\end{equation}
with
\begin{equation}\label{definef}
f(\theta) = \f{(r_+^2 + a^2)[q_e (r_+^2 - a^2\cos^2\t ) + 2q_m ar_+  \cos\t]}{2\rho_+^2 \Xi a r_+}\ .
\end{equation}
This is a generalization of the near horizon extremal Kerr (NHEK) metric
found in \cite{Bardeen:1999px}, and has the same isometries up to a rescaling of $\phi$,
\begin{align}
&K_1 = \p_\phi\ ,\no
&\bar{K}_1 = \p_t\ ,\qquad
\bar{K}_2 = t \p_t - r \p_r, \qquad
\bar{K}_3 = \left({1\over 2r^2} + {t^2\over 2}\right)\p_t - t r \p_r - {k\over r}\p_\phi\ .
\end{align}
These generate $U(1)_L \times SL(2,R)_R$. With the choice of gauge (\ref{nearhorizonA}), $A$ is also invariant under diffeomorphisms generated by $K_1, \bar{K}_{1,2,3}$.

\section{Asymptotic Symmetry Group}\label{s:asg}
The asymptotic symmetry group (ASG) of a spacetime is the group of allowed symmetries modulo trivial symmetries.
A symmetry is allowed if it generates a transformation that obeys the boundary conditions, and it is trivial if its associated charge
(defined below) vanishes.
The definition of the charge associated with a symmetry depends on the action, so the ASG depends on both the action and the choice
of boundary conditions.

In this section we derive the ASG for near horizon extremal 4D black holes with the Einstein-Maxwell action (\ref{einsteinmaxwell}).  It was proved in \cite{KLR,KuLu} that when the cosmological constant is zero or negative, the most general extremal, stationary, rotationally symmetric black hole in this theory has a near horizon metric and gauge field of the form\footnote{
It is assumed that the generator of the rotational symmetry has a fixed point and that the topology of the horizon is not a torus.  The theorem does not apply with positive cosmological constant, but the Kerr-Newman-dS black hole also has this form.}
\begin{eqnarray}\label{asgmetric}
 ds^2&=&\Gamma(\theta)\left[
-r^2dt^2+\frac{dr^2}{r^2}
+\alpha(\theta) d\theta^2\right] + \gamma(\theta)(d\phi+krdt)^2\ , \\
A&=&f(\theta)(d\phi+krdt)\ .\notag
\end{eqnarray}
This class of black holes includes the extremal dyonic  Kerr-Newman black hole as well as its AdS or dS generalization.  Explicit expressions for $f,\Gamma,\gamma,\alpha$, and $k$ in these cases were given in (\ref{definefuncs}, \ref{defineks}, \ref{definef}).
For now, we assume non-zero angular momentum $J > 0$; the $J=0$ (extremal Reissner-Nordstrom) near horizon geometry degenerates to
$AdS_2\times S^2$ and will be considered in section \ref{s:RNBH}.

\subsection{Charges}
To compute the charges associated with asymptotic symmetries, we use the formalism of \cite{barnichbrandt,barnichcompere}.  There is a contribution from the Einstein action as in \cite{GHSS} plus a second contribution from the Maxwell term.  Asymptotic symmetries of (\ref{einsteinmaxwell}) can include diffeomorphisms $\zeta$ under which
\begin{eqnarray}
\delta_\zeta A_\mu &=& \mathcal{L}_\zeta A_\mu\\
\delta_\zeta g_{\mu\nu} &=& \mathcal{L}_\zeta g_{\mu\nu}\notag
\end{eqnarray}
and $U(1)$ gauge transformations $\Lambda$ under which
\begin{equation}
\delta_\Lambda A = d \Lambda\ .
\end{equation}
We denote the infinitesimal field variations by $a_\mu = \delta A_\mu$ and $h_{\mu\nu} = \delta g_{\mu\nu}$.
The combined transformation $(\zeta, \Lambda)$ has an associated charge $Q_{\zeta,\Lambda}$ defined by
\begin{equation}
\delta Q_{\zeta,\Lambda} = {1\over 8 \pi }\int  \left(k_{\zeta}^{grav}[h;g] + k_{\zeta,\Lambda}^{gauge}[h,a; g, A]\right)
\end{equation}
where the integral is over the boundary of a spatial slice.
The contribution from the Einstein action is\footnote{Our sign conventions differ from \cite{GHSS}. We take the coordinate basis $(t,\phi,\t,r)$ and $\int d\phi \wedge d\theta = +1$.}
\begin{eqnarray}\label{chargeintegrand}
k_\zeta^{grav}[h,g]&=& {1\over 4} \epsilon_{\alpha\beta\mu\nu}\big[
\zeta^\nu D^\mu h - \zeta^\nu D_\sigma h^{\mu\sigma} + \zeta_\sigma D^\nu h^{\mu\sigma} \\
& &
\qquad + \half h D^\nu \zeta^\mu - h^{\nu\sigma} D_\sigma \zeta^\mu
+ \half h^{\sigma\nu}(D^\mu\zeta_\sigma + D_\sigma\zeta^\mu)
\big]dx^\alpha\wedge dx^\beta \,.\notag
\end{eqnarray}
The last two terms\footnote{These two terms, and the last two terms in (\ref{kG}), are called the supplementary terms and are absent in the Iyer-Wald formalism \cite{Wa,IyWa}.}
vanish for an exact Killing vector and in most cases do not contribute.
The Maxwell contribution is \cite{barnichbrandt,barnichcompere,comperethesis,barnichcomperegodel,banadoscompere}
\begin{eqnarray}\label{kG}
k_{\zeta,\Lambda}^{gauge}[\delta \phi, \phi] &=& {1\over 8}\epsilon_{\alpha\beta\mu\nu}
\big[(-\half h F^{\mu\nu} + 2F^{\mu\gamma}h_\gamma^{\ \nu} - \delta F^{\mu\nu})(\zeta^\rho A_\rho + \Lambda)\notag \\
& & \qquad - F^{\mu\nu}\zeta^\rho a_\rho - 2 F^{\alpha\mu}\zeta^\nu a_\alpha \big]dx^\alpha \wedge dx^\beta \\
& & - {1\over 8}\epsilon_{\alpha\beta}^{\ \ \mu\nu}a_\mu(\mathcal{L}_\zeta A_\nu + \p_\nu \Lambda)dx^\alpha \wedge dx^\beta \ , \notag
\end{eqnarray}
where $\delta F^{\mu\nu} \equiv g^{\mu\alpha}g^{\nu\beta}(\p_\alpha a_\beta - \p_\beta a_\alpha)$.
Again, the last two terms vanish for an exact symmetry.

The charge $Q_{\zeta, \Lambda}$ generates the symmetry $(\zeta, \Lambda)$ under Dirac brackets.
The algebra of the ASG is the Dirac bracket algebra of the charges themselves,
\begin{eqnarray}\label{algebra}
\{Q_{\zeta,\Lambda}, Q_{\tilde{\zeta}, \tilde{\Lambda}}\}_{DB} &=&
(\delta_{\ti\zeta} + \delta_{\ti\Lambda})Q_{\zeta, \Lambda}\\
&=& {1\over 8 \pi }\int \left(k_{\zeta}^{grav}[\mathcal{L}_{\ti\zeta}g; g] + k_{\zeta,\Lambda}^{gauge}[\mathcal{L}_{\ti\zeta}
g, \mathcal{L}_{\ti\zeta}A + d\ti\Lambda; g, A]\right)\notag\ \\
&=& Q_{[(\zeta,\Lambda), (\tilde\zeta,\tilde\Lambda)]}+{1\over 8 \pi }\int \left(k_{\zeta}^{grav}[\mathcal{L}_{\ti\zeta}\bar g; \bar g] + k_{\zeta,\Lambda}^{gauge}[\mathcal{L}_{\ti\zeta}
\bar g, \mathcal{L}_{\ti\zeta}\bar A + d\ti\Lambda; \bar g, \bar A]\right)\notag\ ,
\end{eqnarray}
where $\bar g, \bar A$ on the last line denote the background solution (\ref{asgmetric}).

\subsection{Boundary Conditions}
For the ASG to be well defined, the $Q_{\zeta,\Lambda}$ must satisfy a number of consistency conditions \cite{barnichbrandt,barnichcompere}.
One condition is that the charges must be finite for all $g, A, h, a$ satisfying the boundary conditions, so
the boundary conditions must be chosen carefully.

For metric fluctuations around the geometry (\ref{asgmetric}), we impose the same boundary conditions used in \cite{GHSS} for the NHEK geometry,
\begin{equation}\label{metricbc}
h_{\mu\nu} \sim \mathcal{O}
\left(
  \begin{array}{cccc}
    r^2 & 1 & 1/r & 1/r^2 \\
     & 1 & 1/r & 1/r \\
     & & 1/r & 1/r^2\\
     &  &  & 1/r^3 \\
  \end{array}
\right)
\end{equation}
in the basis $(t, \phi, \theta, r)$.  Just as for NHEK, an additional nonlinear boundary condition is imposed below to forbid excitations above extremality. This is also needed to render
the charges well defined.  For the gauge field we impose the boundary condition
\begin{equation}\label{abdry}
a_\mu \sim \mathcal{O}(r, 1/r, 1, 1/r^2) \ .
\end{equation}
The most general diffeomorphisms which preserve the boundary conditions on the metric are \cite{GHSS}
\begin{eqnarray}
\zeta_\epsilon &=& \epsilon(\phi)\p_\phi - r \epsilon'(\phi)\p_r\\
\bar\zeta &=& \p_t \notag
\end{eqnarray}
plus subleading terms given in \cite{GHSS}.  We take the basis $\zeta_n$ with $\epsilon_n = -e^{-in\phi}$.
The gauge field transforms under $\zeta_\epsilon$ as
\begin{equation}
\delta_\epsilon A = f \epsilon' (d\phi- k r dt)\ .
\end{equation}
This does not satisfy the boundary condition (\ref{abdry}), so we must add a compensating $U(1)$ gauge transformation to restore
$\delta A_\phi = O(1/r)$.  A similar situation was encountered for the symmetry generators of AdS$_2$ in \cite{HartmanStrominger}.
An alternative might be to loosen the boundary condition to allow $\delta A_\phi = O(1)$, but this does not affect the central charge and one would need to check that such a choice leads to a consistent ASG.
The appropriate compensating gauge transformation is
\begin{equation}
\Lambda = -f(\theta) \epsilon (\phi)\ .
\end{equation}
Under the combined gauge + diffeomorphism transformation,
\begin{equation}
\delta_{\epsilon} A = - k r f(\theta)\epsilon'(\phi) dt - f'(\theta)\epsilon(\phi) d\theta\ .
\end{equation}
The boundary conditions also allow asymptotic gauge transformations
\begin{equation}
\Lambda_a = \Lambda_a(t,\theta) + \mathcal{O}(1/r)\ ,
\end{equation}
where the subleading term can be seen to be trivial after computing the charges.  The leading term is not trivial, but it generates an algebra with no central terms.  Since we wish to focus on the left-moving Virasoro algebra of the extremal black hole, we impose the additional boundary conditions
\begin{equation}\label{extrabc}
Q_{\p_t} = Q_{\Lambda_a} = 0\ .
\end{equation}
As explained in \cite{GHSS}, the restriction to fields satisfying (\ref{extrabc}) is consistent because the generators $\p_t$ and $\Lambda_a$ commute with other generators in the ASG.  In other situations, it may be appropriate to relax (\ref{extrabc}), or perhaps to choose different boundary conditions altogether. With our choice of boundary conditions, the asymptotic symmetries consist of the pairs $(\zeta_n, \Lambda_n)$, with the algebra
\begin{equation}
[(\zeta_n,\Lambda_n), (\zeta_m, \Lambda_m)] = ([\zeta_n,\zeta_m], [\Lambda_n,\Lambda_m]_{\zeta})
\end{equation}
where $[\zeta_n,\zeta_m]$ is the Lie commutator and
\begin{equation}
[\Lambda_n, \Lambda_m]_\zeta = \zeta^\mu_n\p_\mu\Lambda_m - \zeta^\mu_m \p_\mu\Lambda_n\ .
\end{equation}
This is the Virasoro algebra with vanishing central charge,
\begin{equation}
i[(\zeta_n,\Lambda_n), (\zeta_m, \Lambda_m)] = (n-m)(\zeta_{n+m}, \Lambda_{n+m})\ .
\end{equation}

\section{Central Charge}\label{s:centralcharge}
Using (\ref{algebra}) and taking $\Lambda = \Lambda(\theta,\phi)$, the Dirac brackets between symmetry generators are
\begin{multline}\label{genalgebra}
i \{ Q_{\zeta_\epsilon, \Lambda}, Q_{\zeta_{\tilde\epsilon}, \tilde\Lambda}\}_{DB}
=  iQ_{[(\zeta_\epsilon,\Lambda), (\zeta_{\tilde\epsilon},\tilde{\Lambda})]} -{i k\over 16 \pi}\int d\theta d\phi \sqrt{\alpha(\t)\gamma(\t)\over \Gamma(\t)}\biggl(f(\theta)\Lambda\tilde{\epsilon}'
+ \Gamma(\t) \epsilon'\tilde{\epsilon}'' \\ + [f(\theta)^2 + \gamma(\t)]\epsilon\tilde{\epsilon}'
- (\epsilon,\Lambda \leftrightarrow \tilde{\epsilon}, \tilde{\Lambda})\biggl)\ ,
\end{multline}
where the terms including $f(\t)$ comes from the gauge field (\ref{kG}), and the others from the gravitational part (\ref{chargeintegrand}).
The algebra of the ASG is one copy of the Virasoro algebra generated by $(\zeta_n, \Lambda_n)$ with charges $Q_n$.   The boundary conditions (\ref{metricbc},\ref{abdry}) ensure that the $Q_n$ are finite, and from (\ref{genalgebra}) their algebra is
\begin{equation}
i\{Q_m, Q_n\}_{DB} = (m-n)Q_{m+n} + {c \over 12}(m^3 - B m)\delta_{m+n,0}\ ,
\end{equation}
where $B$ is a constant that can be absorbed by a shift in $Q_0$.
The central charge denoted $c_L$ has contributions from $k^{grav}$ and $k^{gauge}$,
\begin{equation}\label{GenCC2}
c  = c_{grav} + c_{gauge}\ .
\end{equation}
We find
\begin{eqnarray}\label{gravgaugec}
c_{grav} &=& 3k \int_0^\pi d\t \s{\G (\t) \a (\t) \g (\t)} \\
c_{gauge} &=& 0 \ .
\end{eqnarray}
This result for $c$ applies to extremal black holes coming from the action (\ref{einsteinmaxwell}).\footnote{More generally, the formula for $c_{grav}$ gives the gravitational contribution to the central charge associated with $\zeta_\epsilon$
for the geometry (\ref{GenExt}).  Black holes with this near horizon metric but in a different theory could have additional
contributions from other terms in the action.}
For the Kerr-Newman-AdS-dS black hole, using (\ref{definefuncs}, \ref{defineks}, \ref{GenCC2}, \ref{gravgaugec}), we find
\begin{align}\label{cTLKNA}
 c  &=\frac{12r_+\sqrt{(3r_+^4/\ell^2 + r_+^2 - q^2)(1-r_+^2/\ell^2 )}}{1+6r_+^2/\ell^2 -3r_+^4/\ell^4 -q^2/\ell^2 }\ .
\end{align}

\section{Temperature}
The Hartle-Hawking vacuum for a Schwarzschild black hole,  restricted to the region outside the horizon, is a density matrix
$\rho=e^{-E/T_H}$ at the Hawking temperature $T_H$. For an extremal black hole, the Hawking temperature vanishes, so one might
expect that the analog of the Hartle-Hawking vacuum (known as the Frolov-Thorne vacuum for Kerr) to be a pure state.
In fact this is not the case because there are additional thermodynamic potentials involved which are conjugate to the charge
and spin.  The extremality constraint requires that any fluctuations satisfy
\begin{equation}
    {0=T_HdS=dM_{ADM}-(\Omega_H dJ +\Phi_e dQ_e+\Phi_mdQ_m ) .}
\end{equation}
where $\Phi_{e,m}$ are electric and magnetic potentials. Equivalently any variation in $J$ or $Q$ is accompanied by an energy variation
\begin{equation}{dM_{ADM}=\Omega_H dJ + \Phi_e dQ_e + \Phi_mdQ_m.}
\end{equation}
For such constrained variations we may write
 \begin{equation}{dS={dJ \over T_L} +{dQ_e \over T_e}+{dQ_m \over T_m} ,}\end{equation}
 where the temperatures are easily computed from the expression for the extremal entropy.
 In the absence of charge or a cosmological constant  we have simply $S=2\pi J$ and $T_L={1 \over 2 \pi}$. More generally
\begin{equation}\label{tempresult}
T_L =\f{(1+6r_+^2/\ell^2 - 3r_+^4/\ell^4 - q^2/\ell^2 ) [2r_+^2(1+r_+^2/\ell^2 ) - q^2]}
{4\pi r_+ [(1+r_+^2/\ell^2 )(1-3r_+^2/\ell^2 ) + q^2/\ell^2 ] \s{(1-r_+^2/\ell^2 )(3r_+^4/\ell^2 + r_+^2 - q^2)}}\ .
\end{equation}
This can also be written $T_L = 1/2\pi k$ where $k$ was defined in (\ref{defineks}).  A similar expression for $T_{e,m}$ will be given when it is needed below.
The generalized Hartle-Hawking vacuum state around an extremal black hole is then the density matrix
\begin{equation}\label{mix}\rho=e^{-{L_0 \over T_L}-{\hat{q}_e \over T_e}-{\hat{q}_m \over T_m}} \end{equation}
where $L_0$ and $\hat{q}_{e,m}$ are the operators for spin and charge.  Since the boundary CFT is dual to the bulk gravity system,
the dual of the black hole is described by the CFT in the mixed state (\ref{mix}).

\section{Entropy}
We have computed the central charge (\ref{cTLKNA}) and temperature (\ref{tempresult}) of the CFT dual to the extreme
Kerr-Newman-AdS-dS black hole. Assuming the Cardy formula, we obtain the statistical entropy of the CFT
\begin{align}\label{cardyf}
S = \f{\pi^2}{3}cT_L
    = \f{\pi(2r_+^4/\ell^2+2r_+^2-q^2)}{1-2r_+^2/\ell^2-3r_+^4/\ell^4+q^2/\ell^2}\ ,
\end{align}
in precise agreement with the Bekenstein-Hawking entropy (\ref{TE}). Note that only the temperature $T_L$ is needed here, as it is the potential conjugate to the zero mode of the Virasoro algebra.

In (\ref{cardyf}) we use the Cardy formula in the canonical ensemble, which is easily derived from the more familiar microcanonical version; for a derivation and related discussion, see \cite{Bousso:2001mw}.  In general, the central charge in the Cardy formula is actually an effective central charge $c_{\mbox{\tiny eff}} = c - 24 \Delta_0$, where $\Delta_0$ is the lowest eigenvalue of $L_0$ \cite{Kutasov:1990sv}.  Here we have assumed that in the semiclassical limit we may take $c\sim c_{\mbox{\tiny eff}}$.  A sufficient but not necessary condition for the validity of the Cardy formula is $T \gg c$.  This condition is obeyed here only for a slowly rotating, highly charged black hole.  In the highly rotating case the condition is violated, but the applicability of the Cardy formula may nonetheless follow from a small mass gap and the existence of highly twisted sectors in the CFT, as discussed for Kerr in \cite{GHSS}.

\section{Reissner-Nordstrom-AdS Black Holes}\label{s:RNBH}
The central charge of the Kerr-Newman-AdS black hole is proportional to $J$.
Therefore, in the limit of the Reissner-Nordstrom-AdS black hole with $J\rightarrow 0$, the central charge approaches zero.
This cancels against the singular behavior of $T_L$ to produce a finite entropy that matches the Bekenstein-Hawking result
\begin{equation}
S_{RN} = \pi r_+^2 \ .
\end{equation}
While the answer matches, the description is clearly singular in this limit.
In this section we propose a dual description of the microscopic entropy of the extremal $J=0$ Reissner-Nordstrom black hole which does not require a singular temperature and central charge.

The electromagnetic field defines an $S^1$ fibered over AdS$_2$.
The $J=0$ description given here requires the additional assumption
that this gauge $S^1$ can be treated as an extra dimension. This allows us to find a Virasoro algebra involving conformal
transformations of the fiber.
While we will see this yields the desired result, it is not clear to us when we expect this to be valid.
In general this means the theory must contain a tower of charged states which correspond to Kaluza-Klein modes and transform into
one another by this Virasoro. In gravity there is always a tower of charged states corresponding to charged black holes.
In string theory,  it is often, but not always, the case that the  gauge $S^1$ can be mapped to a geometric $S^1$ by a duality
transformation, which justifies our assumption. From the worldsheet point of view, a spacetime gauge field implies a worldsheet
current, and the corresponding $U(1)$ worldsheet boson indeed behaves like an extra dimension. Finally, in the case of 5D spinning
black holes, where there is no duality map of the gauge $S^1$ to a geometric one, it was shown in  \cite{GuicaStrominger} that near
maximal spin it nevertheless behaves like a geometric one. So we see the assumption is at least often valid, and we know of no cases where it is not valid.

Turning this around, the success of the black hole  microstate counting based on this assumption suggests that it may always be
valid in a consistent quantum theory of gravity.

The near horizon isometry group of the Kerr-Newman-AdS black hole is only
\begin{equation}
SL(2,R)_R \times U(1)_L
\end{equation}
but there is an additional $U(1)_{gauge}$ symmetry.  We can combine the $U(1)$ gauge bundle with the
geometry and write the 5D total space as \cite{GuicaStrominger}
\begin{equation}\label{totalspace}
ds^2 = ds_{BH}^2 + (dy + A)^2 \ ,
\end{equation}
where $y$ is the fiber coordinate with period $2\pi$ and $ds^2_{BH}$ is the 4D near horizon black hole metric (\ref{NHKNA}).
The geometrical fiber $\phi$ degenerates when the angular momentum vanishes,
$a \rightarrow 0$, and the gauge fiber $y$ degenerates when the charges $q_e, q_m$ vanish.
This is  similar to the 5D Kerr-AdS case studied in \cite{LMP}, with $q_e$ taking the place of the second angular momentum.

Because we are interested in the Reissner-Nordstrom black hole $a=0$, we should choose a gauge for $A$ that is non-singular as
$a \rightarrow 0$.  The simplest choice is
\begin{equation}
A \rightarrow A   -{q_e r_+\over 2a} d\phi
\end{equation}
where the original $A$ was given in (\ref{nearhorizonA},\ref{definef}).
Setting $a = 0$,
\begin{equation}\label{GFRN}
A = q_e r  {\bar{r}_0^2\over r_+^2}dt + q_m \cos\theta d\phi
\end{equation}
where
\begin{equation}\label{definerbar}
\bar{r}_0^2 = \lim_{a\rightarrow 0} r_0^2 = r_+^2 {1 - r_+^2/\ell^2\over 1 + 6 r_+^2/\ell^2 - 3 r_+^4/\ell^4 - q^2/\ell^2}\ .
\end{equation}
\subsection{Central Charge}
Treating the gauge fiber $y$ like a geometric $S^1$ allows us to  extend the $U(1)_{gauge}$
symmetry to a Virasoro algebra generated by
\begin{equation}\label{zetay}
\zeta^{(y)} = \epsilon(y)\p_y  - r \epsilon'(y)\p_r\ .
\end{equation}
To compute the central charge, we treat the 5D total space (\ref{totalspace}) geometrically, so the charges are given by the 5D generalization of (\ref{chargeintegrand}) coming purely from the 5D Einstein action (with Newton's constant $G_N^{(5)} = 2\pi$ in order to reproduce $G_N^{(4)} = 1$ after integrating over $y$).  We choose boundary conditions on the 5D metric
\begin{equation}
h_{\mu\nu} \sim \mathcal{O}\left(
                             \begin{array}{ccccc}
                               r^2 & r & 1/r & 1/r^2 & 1 \\
                                & 1/r & 1 & 1/r & 1 \\
                                &  & 1/r & 1/r^2 & 1/r \\
                                & &  & 1/r^3 & 1/r \\
                                &  &  &  & 1 \\
                             \end{array}
                           \right)\ ,
\end{equation}
in the basis $(t,\phi,\theta,r,y)$.  These are similar to (\ref{metricbc}), but do not allow $\zeta_{\epsilon}$ and do allow $\zeta^{(y)}$.  The most general diffeomorphisms which preserve this boundary condition are of the form
\begin{multline}
\zeta = [b_t + \mathcal{O}(1/r^3)]\p_t + [-r \epsilon'(y) + \mathcal{O}(1)]\p_r + [b_\phi + \mathcal{O}(1/r^2)]\p_\phi \\+ \mathcal{O}(1/r)\p_\theta + [\epsilon(y) + \mathcal{O}(1/r^2)]\p_y\ ,
\end{multline}
where $b_{t,\phi}$ are arbitrary constants.

Following the same steps used to compute the central charge in Kerr/CFT or in section \ref{s:asg}, the central charge associated with $\zeta^{(y)}$ is
\begin{equation}
c_{(y)} = 6 q_e \bar{r}_0^2\ ,
\end{equation}
 where $\bar{r}_0$ was defined in (\ref{definerbar}).

\subsection{Temperature}\label{sbs:TLRN}
The temperature conjugate to electric charge is defined by
\begin{equation}
T_e d S = dQ_e\ ,
\end{equation}
with other charges held fixed.  In the non-rotating case $a=0$,
\begin{equation}\label{RNTL}
T_e = {r_+^2\over 2 \pi q_e \bar{r}_0^2}\ .
\end{equation}

\subsection{Entropy}
The Bekenstein-Hawking entropy of the extremal Reissner-Nordstrom-AdS black hole is
\begin{equation}
S_{BH} = \pi r_+^2\ .
\end{equation}
In the CFT the Cardy formula gives the same result,
\begin{equation}
S_{CFT} = {\pi^2\over 3}c_{(y)} T_e = \pi r_+^2\ .
\end{equation}

\section{More General Black Holes}

Consider the action
\begin{multline}\label{generalaction}
 S=\f{1}{16\pi}\int d^4x\sqrt{-g}\left(
R-\frac{1}{2}f_{AB}(\chi)\partial_\mu \chi^A \partial^\mu \chi^B
-V(\chi)-\frac{1}{4}g_{IJ}(\chi)F^I_{\mu\nu}F^{J\mu\nu}
\right)\\
+\frac{1}{2}\int h_{IJ}(\chi)F^I\wedge F^J\ ,
\end{multline}
where $\chi^A$ are scalar fields, $F^I=dA^I$ are $U(1)$ field strengths, the functions $f_{AB}(\chi)$ and $g_{IJ}(\chi)$ are positive
definite matrices, and the scalar potential is non-positive.  It was shown in \cite{KLR} that if we assume a rotational symmetry with a fixed
point and that the horizon topology is not a torus, then the most general near horizon metric of a stationary, extremal black hole in this theory is of the form (\ref{GenExt}).
The near horizon scalar fields and gauge fields have the form
\begin{equation}\label{gaugebackground}
\chi^A=\chi^A(\theta)\ ,\quad
A^I=f^I(\theta)(d\phi+krdt)\ .
\end{equation}
Here $k$ is constant, $\Gamma$, $\a$, $\gamma$, $\chi^A$ and $f^I$
are unspecified functions of $\theta$, and the coordinates are defined with
$0<\theta<\pi$ and $0<\phi<2\pi$.
The Bekenstein-Hawking entropy of such a black hole is
\begin{equation}\label{generalbekhawk}
    S_{grav} = \f{\pi}{2} \int_0^\pi d\t \s{\G (\t)\a (\t) \g (\t)}\ .
\end{equation}
The action (\ref{generalaction}) also describes certain higher dimensional black holes with $R_t \times U(1)^{D-4}$ symmetry, after Kaluza-Klein reduction and moving to the 4D Einstein frame.  One example is a Kaluza-Klein black hole with internal space $T^{D-4}$.  There are also many five-dimensional black holes with non-trivial horizon topology, including the Myers-Perry black hole, black rings, and black saturns \cite{Myers:1986un,Emparan:2001wn,Pomeransky:2006bd,Elvang:2007rd,Iguchi:2007is,Izumi:2007qx,Elvang:2007hs}, that have $R_t\times U(1)^2$ spacetime symmetry.  Upon Kaluza-Klein reduction to four dimensions, all of these solutions are described by (\ref{generalaction}) and have an additional $U(1)$ isometry, so the statistical entropy of these black holes is also addressed in this section.

It is possible to compute the asymptotic symmetry group of this geometry along the lines of section \ref{s:asg}.
The expression for the asymptotic charges $Q_{\zeta,\Lambda}$ would have contributions from the scalar field in the action
(\ref{generalaction}), and $k^{gauge}$ would be modified by the functions $g_{IJ}(\chi)$ that appear in front of the gauge kinetic term.
Unless there is some obstruction to including $\zeta_{\epsilon}$ in the ASG, the asymptotic symmetries include a Virasoro algebra,
and the gravitational contribution to its central charge was computed in section \ref{s:centralcharge}.

Without first working out the thermodynamics of the general black hole, we cannot compute the temperature of the dual CFT.  However, it is interesting to note that if we naively generalize the formula
\begin{equation}\label{AsTL}
T_L = {1\over 2 \pi k}
\end{equation}
derived for Kerr-Newman-AdS-dS in (\ref{tempresult}), then the Cardy formula gives
\begin{eqnarray}\label{genentrop}
S_{CFT} &=& {\pi^2\over 3}c_{grav}T_L\\
&=& \f{\pi}{2} \int_0^\pi d\t \s{\G (\t)\a (\t) \g (\t)} \notag\\
&=& \f{\text{Area(horizon)}}{4}\ , \notag
\end{eqnarray}
in agreement with (\ref{generalbekhawk}). This prescription to find the temperature of the dual CFT does not depend on the asymptotic structure of the spacetime, so it allows us to reproduce the Bekenstein-Hawking law using only information about the near horizon region.\footnote{For example, in four dimensional Einstein gravity with zero cosmological constant, there is a distorted Schwarzschild black hole that is not asymptotically Minkowski space \cite{FrSa}.  If the action (\ref{generalaction}) has a similar solution with non-trivial asymptotic geometry, then the prescription above will reproduce its entropy.
}  This is plausible because the black hole entropy is an inherent property of the horizon, but (\ref{AsTL}) has not been derived in the general case.

The prescription (\ref{AsTL}) for the temperature works similarly for the Reissner-Nordstrom black hole. From (\ref{totalspace}, \ref{GFRN}), the total space is
\begin{equation}\label{FibreU1}
ds^2 = ds^2_{BH}  + \left( dy + q_e  {\bar{r}_0^2\over r_+^2}rdt + q_m \cos\theta d\phi  \right)^2\ ,
\end{equation}
where $y\sim y + 2\pi$.
According to (\ref{AsTL}), we read off $k=q_e  {\bar{r}_0^2/ r_+^2}$ from the coefficient of $rdt$
and obtain the dual temperature
\begin{equation}
T_y = {r_+^2\over 2\pi q_e \bar{r}_0^2}\ ,
\end{equation}
in agreement with (\ref{RNTL}).

Although the temperature formula (\ref{AsTL}) works empirically, note that we have not accounted for non-gravitational contributions to the central charge in the CFT entropy (\ref{genentrop}).
For the Kerr-Newman-AdS-dS black hole and other black holes with the action (\ref{einsteinmaxwell}), we confirmed in section
\ref{s:centralcharge} that $c_{grav}$ is the only contribution to the central charge; for the more general black holes considered here,
this agreement in the entropy can be taken as evidence that $c = c_{grav}$.
Another possibility is that in the general case,
both the central charge and the temperature receive additional contributions that cancel to produce the correct entropy.

\vspace{1.3cm}
\centerline{\bf Acknowledgements}

We are grateful to M. Guica, H. Irie, C. Keller, S. Minakami and W. Song for valuable discussions, and M. Guica for collaboration on Kerr-Newman at an earlier stage.
TN would like to thank all members of the High Energy Theory Group at Harvard University for
their hospitality during his stay, where important parts of this work were done.
The work of KM is supported by JSPS Grant-in-Aid for Scientific Research
No.\,19$\cdot$3715. The work of TN is supported by JSPS Grant-in-Aid for Scientific Research
No.\,19$\cdot$3589 and the Grant-in-Aid for the Global COE Program
``The Next Generation of Physics, Spun from Universality and Emergence" from the Ministry
of Education, Culture, Sports, Science and Technology (MEXT) of Japan.
The work of TH and AS is supported in part by DOE grant DE-FG02-91ER40654.




\begin{thebibliography}{99}
\bibitem{SV}
  A.~Strominger and C.~Vafa,
  ``Microscopic Origin of the Bekenstein-Hawking Entropy,''
  Phys.\ Lett.\  B {\bf 379}, 99 (1996)
  [arXiv:hep-th/9601029].

\bibitem{Sen}
  A.~Sen,
  ``Black Hole Entropy Function, Attractors and Precision Counting of
  Microstates,''
  arXiv:0708.1270 [hep-th].

\bibitem{GHSS}
  M.~Guica, T.~Hartman, W.~Song and A.~Strominger,
  ``The Kerr/CFT Correspondence,''
  arXiv:0809.4266 [hep-th].

\bibitem{Car}
  S.~Carlip,
  ``What we don't know about BTZ black hole entropy,''
  Class.\ Quant.\ Grav.\  {\bf 15}, 3609 (1998)
  [arXiv:hep-th/9806026].

\bibitem{Car2}
  S.~Carlip,
  ``Black hole entropy from conformal field theory in any dimension,''
  Phys.\ Rev.\ Lett.\  {\bf 82}, 2828 (1999)
  [arXiv:hep-th/9812013].

\bibitem{Park}
  M.~I.~Park,
  ``Hamiltonian dynamics of bounded spacetime and black hole entropy:
  Canonical method,''
  Nucl.\ Phys.\  B {\bf 634}, 339 (2002)
  [arXiv:hep-th/0111224].

\bibitem{KKP}
  G.~Kang, J.~i.~Koga and M.~I.~Park,
  ``Near-horizon conformal symmetry and black hole entropy in any  dimension,''
  Phys.\ Rev.\  D {\bf 70}, 024005 (2004)
  [arXiv:hep-th/0402113].

\bibitem{HHKNT}
  K.~Hotta, Y.~Hyakutake, T.~Kubota, T.~Nishinaka and H.~Tanida,
  ``The CFT-interpolating Black Hole in Three Dimensions,''
  arXiv:0811.0910 [hep-th].

\bibitem{KLR}
  H.~K.~Kunduri, J.~Lucietti and H.~S.~Reall,
  ``Near-horizon symmetries of extremal black holes,''
  Class.\ Quant.\ Grav.\  {\bf 24}, 4169 (2007)
  [arXiv:0705.4214 [hep-th]].

\bibitem{KuLu}
  H.~K.~Kunduri and J.~Lucietti,
  ``A classification of near-horizon geometries of extremal vacuum black
  holes,''
  arXiv:0806.2051 [hep-th].

\bibitem{ANT}
  T.~Azeyanagi, T.~Nishioka and T.~Takayanagi,
  ``Near Extremal Black Hole Entropy as Entanglement Entropy via AdS2/CFT1,''
  Phys.\ Rev.\  D {\bf 77}, 064005 (2008)
  [arXiv:0710.2956 [hep-th]].

\bibitem{RyuTa}
  S.~Ryu and T.~Takayanagi,
  ``Holographic derivation of entanglement entropy from AdS/CFT,''
  Phys.\ Rev.\ Lett.\  {\bf 96}, 181602 (2006)
  [arXiv:hep-th/0603001],

  ``Aspects of holographic entanglement entropy,''
  JHEP {\bf 0608}, 045 (2006)
  [arXiv:hep-th/0605073].

\bibitem{brownhenneaux}
  J.~D.~Brown and M.~Henneaux,
  ``Central Charges in the Canonical Realization of Asymptotic Symmetries: An
  Example from Three-Dimensional Gravity,''
  Commun.\ Math.\ Phys.\  {\bf 104}, 207 (1986).


\bibitem{LMP}
  H.~Lu, J.~Mei and C.~N.~Pope,
  ``Kerr/CFT Correspondence in Diverse Dimensions,''
  arXiv:0811.2225 [hep-th].

\bibitem{AOT}
  T.~Azeyanagi, N.~Ogawa and S.~Terashima,
  ``Holographic Duals of Kaluza-Klein Black Holes,''
  arXiv:0811.4177 [hep-th].

\bibitem{CCK}
  M.~M.~Caldarelli, G.~Cognola and D.~Klemm,
  ``Thermodynamics of Kerr-Newman-AdS black holes and conformal field
  theories,''
  Class.\ Quant.\ Grav.\  {\bf 17}, 399 (2000)
  [arXiv:hep-th/9908022].

\bibitem{Bardeen:1999px}
  J.~M.~Bardeen and G.~T.~Horowitz,
  ``The extreme Kerr throat geometry: A vacuum analog of AdS(2) x S(2),''
  Phys.\ Rev.\  D {\bf 60}, 104030 (1999)
  [arXiv:hep-th/9905099].

\bibitem{Wa}
  R.~M.~Wald,
  ``Black hole entropy is the Noether charge,''
  Phys.\ Rev.\  D {\bf 48}, 3427 (1993)
  [arXiv:gr-qc/9307038].

\bibitem{IyWa}
  V.~Iyer and R.~M.~Wald,
  ``Some properties of Noether charge and a proposal for dynamical black hole
  entropy,''
  Phys.\ Rev.\  D {\bf 50}, 846 (1994)
  [arXiv:gr-qc/9403028].

\bibitem{barnichbrandt}
  G.~Barnich and F.~Brandt,
  ``Covariant theory of asymptotic symmetries, conservation laws and  central
  charges,''
  Nucl.\ Phys.\  B {\bf 633}, 3 (2002)
  [arXiv:hep-th/0111246].

\bibitem{barnichcomperegodel}
  G.~Barnich and G.~Compere,
  ``Conserved charges and thermodynamics of the spinning Goedel black hole,''
  Phys.\ Rev.\ Lett.\  {\bf 95}, 031302 (2005)
  [arXiv:hep-th/0501102].

\bibitem{banadoscompere}
  M.~Banados, G.~Barnich, G.~Compere and A.~Gomberoff,
  ``Three dimensional origin of Goedel spacetimes and black holes,''
  Phys.\ Rev.\  D {\bf 73}, 044006 (2006)
  [arXiv:hep-th/0512105].

\bibitem{barnichcompere}
  G.~Barnich and G.~Compere,
  ``Surface charge algebra in gauge theories and thermodynamic integrability,''
  J.\ Math.\ Phys.\  {\bf 49}, 042901 (2008)
  [arXiv:0708.2378 [gr-qc]].

\bibitem{comperethesis}
  G.~Compere,
  ``Symmetries and conservation laws in Lagrangian gauge theories with
  applications to the mechanics of black holes and to gravity in three
  dimensions,''
  arXiv:0708.3153 [hep-th].

\bibitem{HartmanStrominger}
  T.~Hartman and A.~Strominger,
  ``Central Charge for $AdS_2$ Quantum Gravity,''
  arXiv:0803.3621 [hep-th].

\bibitem{Bousso:2001mw}
  R.~Bousso, A.~Maloney and A.~Strominger,
  ``Conformal vacua and entropy in de Sitter space,''
  Phys.\ Rev.\  D {\bf 65}, 104039 (2002)
  [arXiv:hep-th/0112218].

\bibitem{Kutasov:1990sv}
  D.~Kutasov and N.~Seiberg,
  ``Number Of Degrees Of Freedom, Density Of States And Tachyons In String
  Theory And Cft,''
  Nucl.\ Phys.\  B {\bf 358}, 600 (1991).



\bibitem{GuicaStrominger}
  M.~Guica and A.~Strominger,
  ``Wrapped M2/M5 duality,''
  arXiv:hep-th/0701011.

\bibitem{FrSa}
  V.~P.~Frolov and N.~G.~Sanchez,
  ``Vacuum Energy Density Near Static Distorted Black Holes,''
  Phys.\ Rev.\  D {\bf 33}, 1604 (1986).

\bibitem{Myers:1986un}
  R.~C.~Myers and M.~J.~Perry,
  ``Black Holes In Higher Dimensional Space-Times,''
  Annals Phys.\  {\bf 172}, 304 (1986).

\bibitem{Emparan:2001wn}
  R.~Emparan and H.~S.~Reall,
  ``A rotating black ring in five dimensions,''
  Phys.\ Rev.\ Lett.\  {\bf 88}, 101101 (2002)
  [arXiv:hep-th/0110260].

\bibitem{Pomeransky:2006bd}
  A.~A.~Pomeransky and R.~A.~Sen'kov,
  ``Black ring with two angular momenta,''
  arXiv:hep-th/0612005.

\bibitem{Elvang:2007rd}
  H.~Elvang and P.~Figueras,
  ``Black Saturn,''
  JHEP {\bf 0705}, 050 (2007)
  [arXiv:hep-th/0701035].

\bibitem{Iguchi:2007is}
  H.~Iguchi and T.~Mishima,
  ``Black di-ring and infinite nonuniqueness,''
  Phys.\ Rev.\  D {\bf 75}, 064018 (2007)
  [arXiv:hep-th/0701043].

\bibitem{Izumi:2007qx}
  K.~Izumi,
  ``Orthogonal black di-ring solution,''
  Prog.\ Theor.\ Phys.\  {\bf 119}, 757 (2008)
  [arXiv:0712.0902 [hep-th]].

\bibitem{Elvang:2007hs}
  H.~Elvang and M.~J.~Rodriguez,
  ``Bicycling Black Rings,''
  JHEP {\bf 0804}, 045 (2008)
  [arXiv:0712.2425 [hep-th]].



\end{thebibliography}
\end{document}